# Coulomb Drag in the Exciton Regime in Electron-Hole Bilayers

J. A. Seamons, C. P. Morath, J. L. Reno & M. P. Lilly[*]

*Sandia National Laboratories, Albuquerque, New Mexico 87185, USA.*

We report electrical transport measurements on GaAs/AlGaAs based electron-hole bilayers. These systems are expected to make a transition from a pair of weakly coupled two-dimensional systems to a strongly coupled exciton system as the barrier between the layers is reduced. Once excitons form, phenomena such as Bose-Einstein condensation of excitons could be observed. In our devices, electrons and holes are confined in double quantum wells, and carriers in the devices are induced with top and bottom gates leading to variable density in each layer. Separate contact to each layer allows Coulomb drag transport measurements where current is driven in one layer while voltage is measured in the other. Coulomb drag is sensitive to interlayer coupling and has been predicted to provide a strong signature of exciton condensation. Drag measurement on EHBLs with a 30 nm barrier are consistent with drag between two weakly coupled 2D Fermi systems where the drag decreases as the temperature is reduced. When the barrier is reduced to 20 nm, we observe a consistent increase in the drag resistance as the temperature is reduced. These results indicate the onset of a much stronger coupling between the electrons and holes which leads to exciton formation and possibly phenomena related to exciton condensation.

Two-dimensional (2D) bilayers composed of electrons in one layer and holes in the other are expected to exhibit Bose-Einstein condensation (BEC) of excitons at zero magnetic field[1,2]. Recent progress in several different systems has demonstrated evidence for exciton condensation using both electrical and optical techniques. Quantum Hall bilayer[3] experiments utilize the half-filled Landau level at high magnetic fields to explore BEC in electron-electron[4] and hole-hole[5] bilayers. In optically generated bilayer excitons, evidence for condensation has been building[6,7] and recently BEC has been reported in polariton[8,9] systems. One system where exciton condensation is expected is the electrically generated 2D electron and 2D hole bilayer at zero magnetic field. Here we report evidence for electron-hole pairing in GaAs/AlGaAs based electron-hole bilayers using the Coulomb drag technique. An increase in the drag signal at lower temperature indicates a dramatic increase in coupling between the layers. These results suggest exciton formation at low temperature.

Examples of electron-hole bilayers that behave as 2D Fermi systems have demonstrated for years[10,11,12,13,14,15,16], but creating bilayers with closely spaced layers that can be measured at the low temperatures needed to observe the formation of excitons has been extremely difficult. The devices reported here have narrow barriers between the wells, independent electrical contacts to the electrons and holes, and independently tuneable density in each layer. Equally important to producing the device is employing a measurement that provides a clear means of identifying non-Fermi liquid behaviour. The Coulomb drag measurement[17], where current in one layer induces a voltage in the other, can be quantitatively understood for weakly coupled Fermi liquid bilayers. In this case, the drag resistance develops as a result of interlayer scattering (Coulombic, phonon, etc.) and decreases as the temperature is lowered due to the vanishing phase space for scattering events. At the other extreme, for an exciton condensate the paired electrons and holes want to move together and it is predicted that the drag resistance will increase dramatically at the critical temperature[18] and diverge as the temperature is lowered. Even above the critical temperature for BEC the drag could increase with decreasing temperature due to pairing fluctuations above the critical temperature[19]. This dramatic change of a decreasing drag resistance to an increasing drag resistance is an important theoretical prediction.

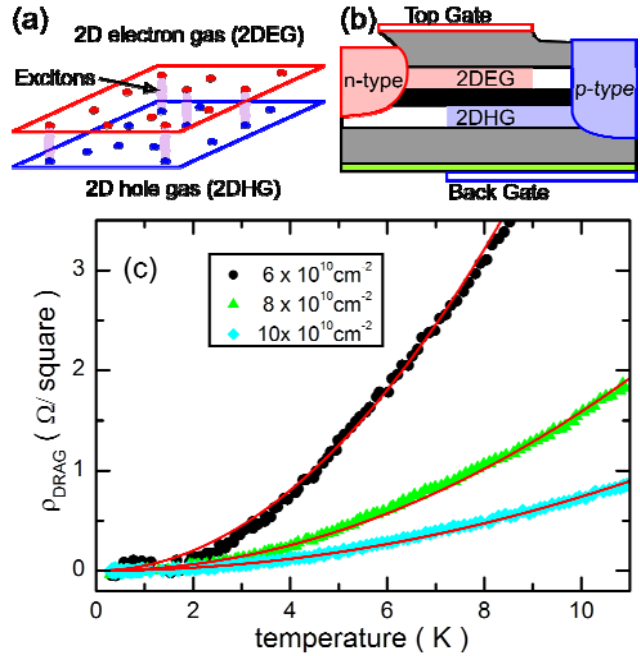

**Figure 1** (color online).   Undoped electron-hole bilayer. (a) Formation of excitons occurs as pairing of electron and hole in opposite layers. (b) Schematic cross section of the uEHBL sample: the conducting areas of the 2DEG (2DHG) are in red (blue), the Al0.3Ga0.7As (Al0.9Ga0.1As) barriers are in grey (black), and a insulating SiN layer is shown in green. (c) Quadratic dependence of drag resistivity for Sample A at matched electron and hole densities of 6 x $10^{10}$ cm$^{-2}$ (black, circles), 8 x $10^{10}$ cm$^{-2}$ (green, triangles), and 1.0 x $10^{11}$ cm$^{-2}$ (cyan, diamonds) is typical for Fermi systems. The red lines are $T^2$ best fits.

The device schematic in Fig. 1b shows the concept of the undoped electron-hole bilayer (uEHBL) GaAs heterostructure.  A detailed explanation of the device fabrication and operation is given in Seamons *et al.*[14].  The 2D electron gas (2DEG) and the 2D hole gas (2DHG) are induced using gates that create an internal electric field in the heterostructure[20,21]. Carriers are pulled into the top (bottom) 18 nm wide GaAs quantum well by applying a voltage between an overall top (bottom) gate and an n-type (p-type) ohmic contacts fabricated in a field effect transistor geometry. For the three samples studied here, the double quantum wells are separated by $Ga_{0.1}Al_{0.9}As$ barriers of either 30 nm for Sample A (wafer EA1286) or 20 nm for Samples B and C (wafer EA1287). This design incorporates both independent contacts and adjustable density of the 2DEG and the 2DHG. To operate the uEHBL a p-type contact of the 2DHG is grounded, while the 2DEG is at a DC interlayer bias $V_{IL}$ ~ -1.45 V needed to overcome the band gap of GaAs; thereby allowing simultaneous occupation of both electrons and holes in these closely spaced layers[22]. Ideally the gates are completely isolated, but in the actual devices we observe leakage currents from the top gate to the n-type contacts and between the electron and hole layers; the leakage currents vary with sample and gate voltage, and do not appear to influence transport measurements. Unfortunately, this leakage between gates does lead to heating at the lowest temperature of a dilution refrigerator, and for the data reported here we used a 3He refrigerator with a base temperature of 0.3 K. Resistance, Hall and Coulomb drag measurements are made using lock-in voltage detection of small AC currents. An isolation transformer is used to combine the AC current through the 2DEG with the required DC voltage $V_{IL}$ of the entire electron layer.

Once the 2DEG and 2DHG are established the density of carriers in each layer is proportional to its respective gate voltage, allowing for independently tuneable densities of the 2DEG (*n*) and the 2DHG (*p*). The individual layer densities were obtained using standard four-terminal longitudinal or Hall resistance at T = 0.3 K. The most accurate *n* and *p* are determined from measuring the Hall slope of each layer independently. At high density, the mobilities exceeded $1 \times 10^6$ cm$^2$/Vs and $4 \times 10^5$ cm$^2$/Vs for the electrons and holes respectively. From the density dependence of the mobility, intra-layer scattering is dominated by background ionized impurities[14,23].

With $n = p$, Coulomb drag measurements were taken by sending a current through the drive layer while measuring the voltage induced in the drag layer. A semi-classical Boltzmann calculation of the Coulomb drag resistivity ($\rho_{DRAG}$) for Fermi systems, assuming high density and large layer spacing[17], reduces to $\rho_{drag} = \alpha T^2 /(np)^{3/2} d^4$ where $\alpha = \hbar \zeta(3)(4\pi\kappa\varepsilon_0)^2 k_B^2 /128\pi e^6$. Here $\hbar$ is Planck's constant, $\zeta(3) \sim 1.202$ is the Riemann zeta function, $\kappa$ is the dielectric constant of GaAs, $\varepsilon_0$ is the permittivity of free space, $k_B$ is Boltzmann's constant, $T$ is the temperature, and $d$ is the center to center distance between the GaAs wells in the drive and drag layers. The results for Sample A where the electrons and holes are spaced by 48 nm are shown in Fig. 1c. All of the drag measurements reported here use $I_{DRIVE} = 50$ nA at 3 Hz in the electron layer. The magnitude of the drive current and the induced voltage in the hole layer are measured simultaneously. As usual for drag measurements, we verified the same drag signal was observed for a range of AC frequencies, varying drive currents in the electron layer, and different ohmic contact configurations. Due to the large contact and sheet resistance of the hole layer, Joule heating at low temperature prevented interchange of the drag and drive layer at low temperature; for T > 1 K, interchange of the drag and drive layers resulted in the same drag resistivity. Using the Shubnikov-de Haas (SdH) minima in the electron layer, we directly measured the heating caused by driving current in the hole layer. Without current in the hole layer, the SdH minima in the electron layer continue to decrease all the way to $T = 0.3$ K. When current is driven in the hole layer, we observe a saturation of the temperature dependence for both SdH minima and Coulomb drag at $T \sim 1$ K. While lowering the current reduces this effect, the small drag signals become difficult to measure.

As expected for Coulomb scattering in Fermi systems, the drag is approximately proportional to $T^2$ and lower densities result in larger $\rho_{DRAG}$. The red lines in Fig. 1c are $T^2$ best fits to $\rho_{DRAG}$ over the range from $T = 0.3$ K to 10 K. Quantitatively, the prefactor $\alpha$ is a factor of 7 to 10 larger than the approximate expression for drag noted above. Such an enhancement has been observed for hole-hole bilayer[24,25,26] and electron-hole bilayers[10], and can be understood in part by deviations from the large layer spacing and high density limit where the $T^2$ approximation is valid. More detailed scattering calculations including realistic modelling of the actual device structure are in good quantitative agreement with the data[23]. The qualitative agreement of the drag resistivity in Sample A with the scattering theory for drag indicates widely spaced layers behave as independent Fermi-liquid systems down to a temperature of $T = 0.3$ K.

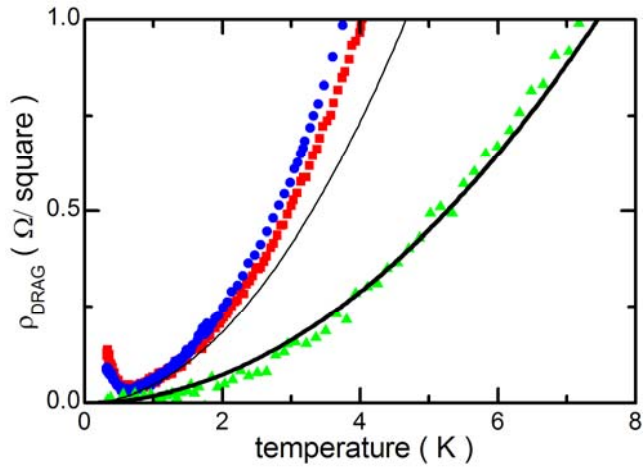

**Figure 2** (color online). $\rho_{DRAG}$ as a function of temperature at $n = p = 8 \times 10^{10}$ cm$^{-2}$ for all three devices. Sample A (30 nm barrier) is plotted with green triangles, Sample B (20 nm barrier) with red squares, and Sample C (20 nm barrier) with blue circles. The thick line is a $T^2$ best fit for Sample A. The thin line is obtained by multiplying the thick line by the ratio $(d_{30\,nm\sim barrier}/d_{20\,nm\sim barrier})^4$.

The most significant result reported here is found when Coulomb drag is measured in devices where the electrons and holes are closer together. In Fig. 2, the drag resistance for all three devices is shown for fixed densities of $n = p = 8 \times 10^{10}$ cm$^{-2}$. A $T^2$ best fit line for data taken on Sample A is shown with a thick black line. Multiplying the thick line best fit for Sample A by $(48 \text{ nm} / 38 \text{ nm})^4$ yields the thin black line. The good agreement between the lines in Fig. 2 and the data above 0.5 K indicates the 2D electrons and holes behave as Fermi liquids, but below 0.5 K in Sample B and Sample C a significant deviation develops. At a critical temperature ($T_u$) the drag reaches a minimum, and for lower temperatures there is a pronounced *upturn* of the drag where $\rho_{DRAG}$ increases with decreasing temperature. Measuring $\rho_{DRAG}$ over different regions of the Hall bar, or with reversed source and drain contacts for the current yields the same results shown in Fig. 2. Further reduction of the temperature in a dilution refrigerator (not shown here) resulted in a saturation of the drag resistivity below T = 0.2 K. Anomalous drag with non-monotonic temperature dependence has been recently reported in Ref. 16.

The observation of an increasing drag resistance at low temperature is a clear and qualitative deviation from expectations based on conventional interlayer scattering theory for Coulomb drag. Before considering exciton formation, we first address a number of recent results that could lead to unusual low temperature drag results. First, theoretical calculations of Coulomb drag calculating to the third order for interlayer interactions predict a finite drag resistance at zero temperature[27]. The predicted magnitude of the finite drag for parameters appropriate to the data in Fig. 2 is $10^{-5}$ $\Omega/\square$ which is far smaller than $0.1$ $\Omega/\square$ observed at low temperature. It is unlikely that third order effects are responsible for the behavior of the drag at low temperature. The second possibility is related to fluctuations in the drag resistivity at zero magnetic field[28] where positive and negative swings on the order of $10^{-2}$ $\Omega/\square$ are observed with changes in density and temperature. In our results with 200 μm wide Hall bars, the upturn is very repeatable for different devices and cooldowns. We do not observe these mesoscopic fluctuations of the drag resistivity. The third possibility we consider is an increase in the coupling between the electron and hole layers as exciton formation occurs.

Exciton formation can affect the drag in a number of ways. One possibility is that $T_u$ corresponds to the critical temperature below which an exciton condensate forms. Vignale and MacDonald[18] predict a discontinuity of the drag at the critical temperature where excitons form a condensate and then a subsequent divergence of the drag as $T \to 0$. In our data, we do not observe a discontinuity, but rather a more gradual change in the slope. The experimental system does have disorder, density variations, finite currents used in measurements, complicated hole bandstructure and many other details that could complicate understanding of any transition to an exciton condensate. Another possibility is that excitons form and condense above the critical temperature for a brief time before rejoining the Fermi sea. This was studied by Hu[19] and is similar to predictions of pairing of composite fermions[29] related to experimental results showing a saturation and increase of the drag at high magnetic field.[30] With a number of approximations the drag was found to increase as log(*T*) as the temperature approaches the critical temperature from above. Due to the very narrow range of temperature where the drag increases, we were not able determine an obvious functional form of the data; in fact, we could fit to both a logarithmic increase and an activated increase in drag. Qualitatively, however, the results for pairing fluctuations have similar trends to the measured Coulomb drag. Finally, another possibility that would result in an enhanced drag signal is exciton formation at temperatures too high for condensation to occur. In any of these scenarios the upturn in drag signals the formation of interlayer electron-hole pairs at zero magnetic field.

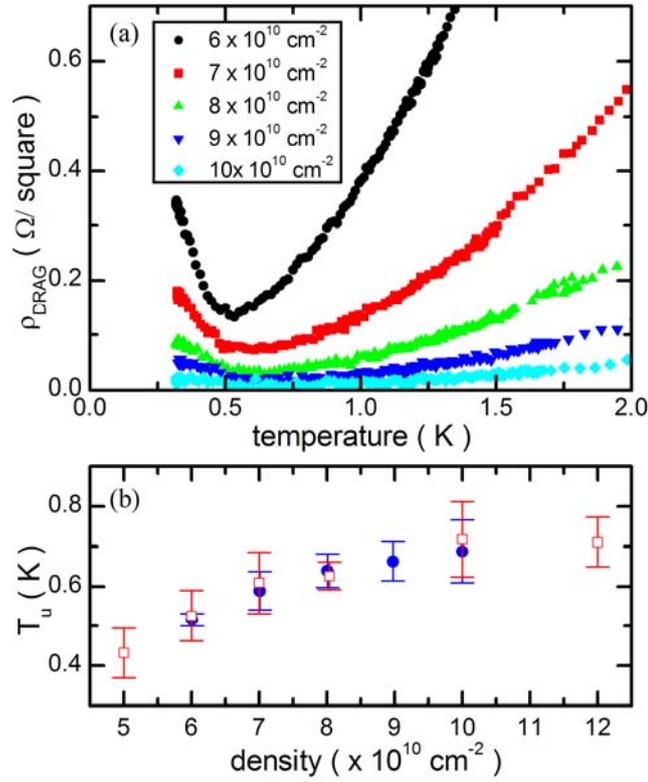

**Figure 3** (color online). (a) $\rho_{DRAG}$ of Sample C (20 nm barrier) as a function of temperature for $n = p = 6 \times 10^{10}$ cm$^{-2}$ (black circles) to $1.0 \times 10^{11}$ cm$^{-2}$ (cyan diamonds) in steps of $1 \times 10^{10}$ cm$^{-2}$. (b) $T_u$ as a function of $n = p$ for Sample B (red open squares) and Sample C (blue circles), with appropriate error bars.

The $\rho_{DRAG}$ at T = 0.3 K in Sample C was also measured as a function of perpendicular magnetic field, without any dramatic change in the magnitude of the effect. The $\rho_{DRAG}$ upturn was independent of the magnetic field up to B = 0.5 T, at which point the drag signal exhibits magnetoresistance oscillations due to the reduced density of states between Landau levels. The upturn in $\rho_{DRAG}$ was observed in two different samples (Sample B and Sample C) that have leakage paths which are substantially different from each other. This indicates that the upturn is not simply an anomaly arising from device leakage. The upturn does not exhibit a maximum value at balanced densities n = p and for unbalanced densities the temperature dependence continues to exhibit an upturn even when the densities differ by up to 30%. Finally, in the density regime where these measurements have been made, neither the electron resistivity or hole resistivity exhibits insulating behaviour.

In Fig. 3a $\rho_{DRAG}$ of Sample C is plotted as a function of temperature for five matched densities cases ranging from $6 \times 10^{10}$ cm$^{-2}$ to $1.0 \times 10^{11}$ cm$^{-2}$. The magnitude of $\rho_{DRAG}$ at a given temperature is strongly dependent on the density. It is evident in Fig. 3a that a minimum in $\rho_{DRAG}$ at $T_u$ is identifiable for each density case in this range. While the absolute magnitude of the upturn is largest for the lowest density signal, there is clearly an upturn at every density studied even though the overall drag at high density is reduced. Fig. 3b shows $T_u$ with approximate error bars associated for Sample B and Sample C as a function of their matched densities. It is interesting to note that at high density where drag signals are small, $T_u$ occurs at higher temperature.

In conclusion, temperature dependent Coulomb drag measurements are presented as a function of matched densities for electron hole bilayers with two different center to center separation. For the larger barrier, the drag resistance can be well described by interlayer Coulomb scattering between fermions. For the narrow barrier device, we observe an increase in the Coulomb drag as the temperature is lowered. The upturn in $\rho_{DRAG}$ was observed in two samples its presence demonstrates increased coupling between the layers. While it is difficult to demonstrate coherence with a Coulomb drag measurement, the increased coupling suggests pairing of electrons and holes and the formation of excitons in electrically generated electron-hole bilayers.

This work has been supported by the Division of Materials Sciences and Engineering, Office of Basic Energy Sciences, U.S. Department of Energy. Sandia is a multiprogram laboratory operated by Sandia Corporation, a Lockheed Martin Company, for the United States Department of Energy under Contract No. DE-AC04-94AL85000. The authors are grateful to J. Eisenstein, S. Das Sarma, A. Balatsky, A. MacDonald, B.

Y. K. Hu, E. H. Hwang and P. Littlewood for discussions and D. R. Tibbetts for fabrication support.

*Correspondence should be addressed to M.P.L. (mplilly@sandia.gov)
[1] S. I. Shevchenko, Sov. J. Low Temp. Phys. **2,** 251 (1976).

[2] Y. E. Lozovik and V. I. Yudson, Sol. St. Commun. **19,** 391 (1976).

[3] J. P. Eisenstein and A. H. MacDonald, Nature **432,** 691 (2004).

[4] M. Kellogg, J. P. Eisenstein, L. N. Pfeiffer and K. W. West, Phys. Rev. Lett. **90,** 246801 (2003).

[5] E. Tutuc, M. Shayegan and D. A. Huse, Phys. Rev. Lett. **93,** 036802 (2004).

[6] L. V. Butov, A. Zrenner, G. Abstreiter, G. Bohm and G. Weimann, Phys. Rev. Lett. **73,** 304 (1994).

[7] L. V. Butov, C. W. Lai, A. L. Ivanov, A. C. Gossard and D. S. Chemla, Nature **417,** 47 (2002).

[8] J. Kasprzak, M. Richard, S. Kundermann, A. Baas, P. Jeambrun, J. M. J. Keeling, F. M. Marchetti, M.. H. Szymanska, R. Andre, J. L. Staehli, V. Savona, P. B. Littlewood, B. Deveaud and L. S. Dang, Nature **443,** 409 (2006).

[9] R. Balili, V. Hartwell, D. Snoke, L. Pfeiffer and K. West, Science **316,** 1007 (2007).

[10] U. Sivan, P. M. Solomon and H. Shtrikman, Phys. Rev. Lett. **68,** 1196 (1992).

[11] H. Rubel, A. Fischer, W. Dietsche, K. v. Klitzing and K. Eberl, Mater. Sci. Eng. B **51,** 207 (1998).

[12] M. Pohlt, M. Lynass, J. G. S. Lok, W. Dietsche, K. v. Klitzing, K. Eberl and R. Muhle, Appl. Phys. Lett. **80,** 2107 (2002).

[13] J. A. Keogh, K. D. Gupta, H. E. Beere, D. A. Ritchie and M. Pepper, Appl. Phys. Lett. **87,** 202104 (2005).

[14] J. A. Seamons, D. R. Tibbetts, J. L. Reno and M. P. Lilly, Appl. Phys. Lett. **90,** 052103 (2007).

[15] K. Das Gupta, M. Thangaraj, A. F. Croxall, H E. Beere, C. A. Nicoll, D. A. Ritchie and M. Pepper, Physica E **40**, 1693 (2008).

[16] A. F. Croxall, K. Das Gupta, C. A. Nicoll, M. Thangaraj, H. E. Beere, D. A. Ritchie and M. Pepper, arXiv:0807.0134 [cond-mat.mes-hall].

[17] T. J. Gramila, J. P. Eisenstein, A. H. MacDonald, L. N. Pfeiffer and K. W. West, Phys. Rev. Lett. **66,** 1216 (1991).

[18] G. Vignale and A. H. MacDonald, Phys. Rev. Lett. **76,** 2786 (1996).

[19] B. Y. K. Hu, Phys. Rev. Lett. **85,** 820 (2000).

[20] B. E. Kane, L. N. Pfeiffer, K. W. West and C. K. Harnett, Appl. Phys. Lett. **63,** 2132 (1993).

[21] M. P. Lilly, J. L. Reno, J. A. Simmons, I. B. Spielman, J. P. Eisenstein, L. N. Pfeiffer, K. W. West, E. H. Hwang and S. Das Sarma, Phys. Rev. Lett. **90,** 56806 (2003).

[22] A more detailed study of the role of interlayer bias in uEHBL devices can be found in C. P. Morath, J. A. Seamons, J. L. Reno and M. P. Lilly, arXiv:0803.1402 [cond-mat.mes-hall].

[23] E. H. Hwang and S. Das Sarma, arXiv:0804.3311 [cond-mat.mes-hall].

[24] R. Pillarisetty, H. Noh, D. C. Tsui, E. P. D. Poortere, E. Tutuc, and M. Shayegan, Phys. Rev. Lett. **89,** 016805 (2002).

[25] E. H. Hwang, S. Das Sarma, V. Braude and A. Stern, Phys. Rev. Lett. **90,** 086801 (2003).

[26] S. Das Sarma and E. H. Hwang, Phys. Rev. B **71,** 195322 (2005).

[27] A. Levchenko and A. Kamenev, Phys. Rev. Lett. **100**, 026805 (2008).

[28] A. S. Price, A. K. Savchenko, B. N. Narozhny, G. Allison and D. A. Ritchie, Science **316**, 99 (2007).

[29] I. Ussishkin and A. Stern, Phys. Rev. Lett. **81**, 3932 (1998).

[30] M. P. Lilly, J. P. Eisenstein, L. N. Pfeiffer and K. W. West, Phys. Rev. Lett. **80**, 1714 (1998).